\documentclass[11pt,oneside]{amsart}
\newcommand{\noi}{\vspace{12pt}\noindent}

\def\beq{\begin{equation}}
\def\eeq{\end{equation}}
\newcommand{\bea}{\begin{eqnarray}}
\newcommand{\eea}{\end{eqnarray}}
\newcommand{\gh}[1]{{\rm gh}(#1)}
\newcommand{\p}[1]{{\epsilon}(#1)}
\def\half{\frac{1}{2}}
\def\manM{{\mathcal M}}
\newcommand{\lpart}{\raise.3ex\hbox{$\stackrel{\leftarrow}{\partial}$}}
\newcommand{\rpart}{\raise.3ex\hbox{$\stackrel{\rightarrow}{\partial}$}}
\newcommand{\ldr}{\raise.3ex\hbox{$\stackrel{\leftarrow}{\delta^r}$}}
\newcommand{\deder}[1]{{ 
 {\stackrel{\raise.1ex\hbox{$\leftarrow$}}{\delta^r}   } 
\over {   \delta {#1}}  }}
\newcommand{\dedel}[1]{{ 
 {\stackrel{\lower.3ex \hbox{$\rightarrow$}}{\delta^l}   }
 \over {   \delta {#1}}  }}

\newcommand{\papar}[1]{{ 
 {\stackrel{\raise.1ex\hbox{$\leftarrow$}}{\partial^r}   } 
\over {   \partial {#1}}  }}
\newcommand{\papal}[1]{{ 
 {\stackrel{\lower.3ex \hbox{$\rightarrow$}}{\partial^l}   }
 \over {   \partial {#1}}  }}

\newcommand{\der}{{\mathcal{D}}}
\newcommand{\cP}{{\mathcal{P}}}

\newcommand{\dl}[1]{\frac{{\partial}}{\partial #1}}

\newcommand{\vardr}[1]{\frac
{{\stackrel{\leftarrow}{\delta}}}{ \delta #1}}
\newcommand{\vardl}[1]{\frac
{{\delta}}{\delta #1}}

\newcommand{\pb}[2]{\left\{{}#1{},{}#2{}\right\}}
\newcommand{\pg}[2]{\left[{}#1{},{}#2{}\right]}
\newcommand{\ab}[2]{\left({}#1{},{}#2{}\right)}

\def\cP{{\mathcal P}}

\def\cE{{\mathcal E}}

\def\manM{{\mathcal M}}

\begin{document}
\thispagestyle{empty}
\vspace{3cm}
\title{{\bf Superfield BRST Charge and the Master Action }}
\vspace{2cm}
\author{{\sc M.A. Grigoriev}\\
Lebedev Physics Institute\\
53 Leninisky Prospect\\Moscow 117924\\Russia\\~\\and\\~\\
{\sc P.H.~Damgaard}\\The Niels Bohr Institute\\Blegdamsvej 17\\
DK-2100 Copenhagen\\Denmark}
\begin{abstract}
Using a superfield formulation of extended phase space, we propose a
new form of the Hamiltonian action functional. A remarkable feature of
this construction is that it directly leads to the BV master action
on phase space. Conversely, superspace can be used to construct nilpotent
BRST charges directly from solutions to the classical Lagrangian
Master Equation. We comment on the relation between these constructions
and the specific master action proposal of Alexandrov, Kontsevich, Schwarz 
and Zaboronsky. 
\end{abstract}
\maketitle
\vspace{10cm}
\begin{flushleft}
NBI-HE-99-46\\hep-th/9911092
\end{flushleft}

\vfill
\newpage

\section{Introduction}

\noi
In two recent papers \cite{super,super1}, it has been shown that quantization, 
in both the Hamiltonian operator language and a phase space path integral, 
has an equivalent superfield formulation.  The superspace consists of 
ordinary time $t$ and a new Grassmann-odd direction $\theta$.  All original 
phase space coordinates $z^A_0(t)$ are just the zero-components of
super phase-space coordinates
\begin{equation}
z^A(t,\theta) ~=~ z^A_0(t) + \theta z^A_1(t) ~.
\label{z}
\end{equation}
It follows that $z^A(t,\theta)$ has the same statistics as $z^A_0(t)$,
denoted by $\p{z^A}$.  The superspace derivative 
\begin{equation}
D ~\equiv~ \frac{d}{d\theta} + \theta \frac{d}{dt} ~,
\label{D}
\end{equation}
replaces the ordinary time derivative.  It satisfies
\begin{equation}
D^2 ~=~ \frac{d}{dt} ~,
\end{equation}
and one can indeed show that the usual Heisenberg equations of motion on the
original phase space variables are obtained by applying $D$ twice on the
extended variables.

\noi
We shall here consider another extension of the usual time derivative,
\begin{equation}
\der ~\equiv~ \theta\frac{d}{dt} ~,
\end{equation}
which obviously satisfies $\der^2=0$.  This derivative will turn out
to play a central role in understanding BV-quantization \cite{BV} in
phase space \cite{FH}.  
We note that $\der$ can be assigned a definite ghost number
according to the ghost number we assign to $\theta$, and we choose
gh($\theta) = 1$.

\noi
We now make some general observations concerning a relation between
the even Poisson bracket on the original phase-space manifold and an 
associated (Grassmann-odd) antibracket on the super path space that
can be induced by it~\cite{super,[AKSZ]}:\footnote{See section 
\ref{sec:geometry} for details.}
\begin{equation}
\begin{array}{rcl}
\label{eq:AB}
\ab{z^A_0(t)}{z^B_0(t^\prime)} &=& 0 \cr
\ab{z^A_0(t)}{z^B_1(t^\prime)} &=& \omega^{AB}(z_0) \delta(t-t^\prime) \cr
\ab{z^A_1(t)}{z^B_1(t^\prime)} &=&
z^C_1\partial_C
 \omega^{AB}(z_0) \delta(t-t^\prime)\,.
\end{array}
\end{equation}
where $\omega^{AB}(z)=\pb{z^A}{z^B}$.  This antibracket obviously
carries one unit of Grassmann parity and one unit of ghost number.  Another
important property of~\eqref{eq:AB} is the following: let $f(z)$ be a
function on phase space and let the functional $F$ be defined by
\begin{equation}
\label{eq:association1}
F[z] ~=~ \int\! dt d\theta ~f(z(t,\theta)) ~,
\end{equation} 
so that the Grassmann parity of $F$ is opposite that of $f$.
Then for any functions $f,g$ on the phase space we have
\begin{equation}
(F,G) ~\equiv~ \int\! dt d\theta~ \{f,g\} ~.\label{eq:ABmain}
\end{equation}
where $F,G$ are the corresponding functionals
obtained by~\eqref{eq:association1}.  In particular, if a
Grassmann-odd function $f$ satisfies the ``Hamiltonian 
Master Equation'' $\{f,f\}=0$ then the corresponding Grassmann-even 
functional $F=\int dtd\theta\, f$ satisfies the BV Master Equation 
with respect to the antibracket:
\begin{equation}
  (F,F)=0\,.
\end{equation}
This has an obvious generalization, because even if $f=f(z,t,\theta)$ 
is a function on the original phase space that also explicitly depends
on time $t$ and its super-partner $\theta$, and $f$ still satisfies the
Hamiltonian Master Equation $\{f,f\}=0$ for any $t$ and $\theta$, then
the corresponding functional $F=\int dt d\theta f(z(t,\theta),t,\theta)$
also satisfies the Master Equation with respect to the antibracket
(\ref{eq:ABmain}).

\section{Superfield Realization of the Phase Space Antifield Formalism} 

\noi

Let us now consider a system with first class constraints.  
We thus have a Grassmann-odd BRST generator
$\Omega = \Omega(z)$ and an 
Hamiltonian $H = H(z)$, which are taken to satisfy
\begin{equation}
\{\Omega,\Omega\} ~ = ~  0
 \qquad {\mbox{\rm and}} \qquad \{H,\Omega\} ~ = ~ 0\,.
\label{homegapb}
\end{equation}
They combine nicely into one Grassmann-odd object $Q$ \cite{super}
\begin{equation}
Q(z,\theta) ~ \equiv ~ \Omega(z) + 
\theta H(z) ~,
\end{equation}
which is nilpotent due to eq. (\ref{homegapb}):
\begin{equation}
\label{qpb}
\{Q,Q\} ~=~  0 \,.
\end{equation}

\noi
Let us in addition consider the following action functional:
\begin{equation}
\label{S}
S[z] ~=~ \int \!dtd\theta ~[V_A(z(t,\theta))\der 
z^A(t,\theta) - Q]\,,
\end{equation}
where the symplectic potential $V_A$ is related to the symplectic
metric $\omega_{AB}$ ($\omega_{AC}\omega^{CB}=\delta_A^B$) via
\begin{equation}
\label{rot}
\omega_{AB} ~=~ \left(\partial_AV_B - (-1)^{\p{A}\p{B}}
\partial_BV_A\right)(-1)^{\epsilon_{B}} \,.
\end{equation}
As usual, we assume the symplectic form to be exact.
We emphasize that this construction is quite different from that of
refs.~\cite{super,super1}. In particular, in the present
formulation the superpartners $z_1(t)$ cannot be viewed as Pfaffian
ghosts. To construct the path integral one must thus in addition specify
the formally invariant super Liouville measure, as usual.
If in addition to eq. (\ref{qpb}) we also assume that there are no boundary
terms from $Q$ 
\begin{equation}
\int\! dtd\theta~\der Q 
~=~ \int \! dt~ \dot{\Omega}(z_0) ~=~ 0~,
\end{equation}
then it follows from the previous considerations that this action satisfies
the Master Equation
\begin{equation}
(S,S) ~=~ 0 ~.
\end{equation}
It is simple to integrate out the additional $\theta$-variable from eq.
(\ref{S}), and one finds:
\begin{equation}
\label{eq:FHmaster}
S[z] ~=~ \int \!dt ~[V_A(z_0(t))
\dot{z}_0^A(t) - H(z_0(t)) - z_1^A(t)\partial_A
\Omega(z_0(t))] \,.
\end{equation}
Except for the last term, this is simply the conventional phase-space
action if there were no constraints.  Let us rewrite this last term:
\begin{equation}
z_1^A\partial_A\Omega(z_0(t)) ~=~ z^*_{0A}\{z_0^A,\Omega\} ~,
\end{equation}
where we have defined $z^*_{0A} \equiv z_1^B\omega_{BA}$.  This is the general
phase space action extended with antifields to satisfy the classical
BV Master Equation $(S,S)=0$. Note that the variables $z^A_0$ and $z^*_{0A}$
are canonically conjugate within the antibracket, but the precise 
identification of which plays the role of ``field'' or ``antifield'' 
becomes apparent upon identification of ghost number \cite{FH}. All these
assignments follow automatically from the superfield approach.

\noi
To make these considerations more concrete, let us consider the case of
first class constraints $T_{\alpha}(z)$ with the usual algebra
\begin{equation}
\{T_\alpha,T_\beta\} ~=~ C^\gamma_{\alpha \beta}T_{\gamma}
\,,\qquad
\{T_\alpha,H_0\} ~=~ V^\beta_\alpha T_\beta \,.
\end{equation}
where for simplicity we assume all the original phase space variables
to be bosonic.  According to the BFV prescription \cite{BFV} one introduces
ghosts $c^{\alpha}$ together with their conjugate momenta $\cP_{\alpha}$, 
and the BRST charge and the extended Hamiltonian are then given by
\bea
  \label{eq:Omega-and-H}
\Omega &=& T_\alpha c^\alpha-
\half \cP_\gamma C^\gamma_{\alpha\beta}c^\alpha c^\beta + \ldots \cr
H &=& H_0 + \cP_\alpha V^\alpha_\beta c^\beta + \ldots
\eea
where the dots denote the higher order terms in the expansion
of $\Omega$ and $H$ with respect to the ghost momenta $\cP$.
The extended Poisson bracket is given in coordinates by
\begin{equation}
  \pb{q^i}{p_j} = \delta^i_j \,, \qquad
\pb{c^\alpha}{\cP_\beta}=\delta^\alpha_\beta\,.
\end{equation}

\noi
We now allow all the phase space coordinates to depend on $t$ and $\theta$.
Let us write explicitly their expansion with respect to $\theta$:
\begin{equation}
\begin{array}{rclrcl}
  \label{eq:exp}
q^i & =& q_0^i - \theta p_*^i\, \quad & \quad
p_i & = & p_i^0 + \theta q_i^* \cr
c^\alpha & = & c_0^\alpha - \theta u^\alpha \,\quad & \quad 
\cP_\alpha & = & u^*_\alpha+\theta c^*_\alpha \,.
\end{array}
\end{equation}
where we have chosen some defining signs in order to facilitate comparison
with the existing literature.
According to our choice $\gh{\theta}=1$ the ghost numbers of the new
variables read
\begin{equation}
\label{eq:ghost}
\begin{array}{llll}
 \gh{q_0^i}=0\,,& 
 \gh{p^0_i}=0\,,&
  \gh{q^*_i}=-1\,,&
  \gh{p_*^i}=-1\,, \cr
  \gh{c_0^\alpha}=1\,,&
  \gh{u^\alpha}=0\,, &
  \gh{c^*_\alpha}=-2\,, &
  \gh{u^*_\alpha}=-1\,.
\end{array}
\end{equation}
The action \eqref{eq:FHmaster} then takes the form
\begin{equation}
\begin{array}{ccl}
\label{eq:STexplicit}
S &=& \int\! dt ~[
p^0_i\dot{q}_0^i - H_0 + T_\alpha u^\alpha-
p_*^i \{p^0_i,T_\alpha\}c^\alpha_0\cr
&&-q^*_i \{q_0^i,T_\alpha\}c^\alpha_0 +
u^*\dot{c}^\alpha_0
-u^*_\alpha V^\alpha_\beta c_0^\beta
+\half c^*_\gamma C^\gamma_{\alpha \beta}c_0^\alpha c_0^\beta-
u^*_\gamma C^\gamma_{\alpha \beta}u^\alpha c_0^\beta+ \ldots] ~,
\end{array}
\end{equation}
where the dots denote higher order terms in powers of
$u^*_\alpha$ and $c^*_\alpha$.
It is easy to see that the first three
terms in (\ref{eq:STexplicit}) are nothing but the extended Hamiltonian action
\begin{equation}
  \label{eq:Eaction}
  S ~=~ \int\! dt~ (p^0_i\dot{q}_0^i - H_0 + T_\alpha u^{\alpha})\,,
\end{equation}
corresponding to
the system under consideration. Indeed, these terms enter only with ghost
number zero variables, and should thus be understood
in the BV formalism as the initial action. Making use of
the ghost number assignments~\eqref{eq:ghost} it is also easy to
infer the gauge generators from eq. (\ref{eq:STexplicit}): they are
precisely the gauge generators of the extended Hamiltonian
action~\eqref{eq:Eaction}. It follows that $u^\alpha$ are simply the
Lagrangian multipliers corresponding to the constraints
$T_\alpha$. All other assignments coincide exactly with those of the extended 
phase space action first identified by Fisch and Henneaux \cite{FH}.  
We have thus explicitly confirmed the remarkable fact
that the whole extended phase-space BV formalism is precisely encoded
in this superspace path integral approach.

\section{An Inverse Construction}\label{sec:invers}

\noi
There are two equivalent ways two perform path integral
quantization of Hamiltonian systems with first-class constraints:
\begin{itemize}
\item[(i)]
via the BFV formalism based on an extended Poisson bracket
and BRST charge $\Omega$.
\item[(ii)]
via the BV formalism based on the antibracket
and the master action corresponding to
the extended Hamiltonian action. 
\end{itemize}
As we have shown above, 
the superfield approach allows one to explicitly derive the
BV formulation from the BFV prescription on phase space. The space of field 
histories (which is the BV configuration space) thus appears as the space of 
super-paths of the initial BFV phase space.  Remarkably, this space comes 
with a BV antibracket which originates directly from the BFV 
Poisson bracket. Similarly, the master action derives directly
from the BRST charge $\Omega$ and the BFV extended Hamiltonian.

\noi
It is natural to ask if there, conversely, exists a ``phase space''
description of any Lagrangian gauge theory which is dual to the standard 
BV description.  As we shall now show, the answer to this is affirmative. 
Moreover, we will again directly arrive at the correct dual description 
by means of the superfield approach. A quite different superfield formulation
of BV Lagrangian quantization was first proposed in ref. \cite{L}.

\noi
Let us start with the standard BV formulation of any Lagrangian
gauge theory. Let $\manM$ be the antisymplectic manifold of the BV 
configuration space, the antisymplectic structure of which determines the
BV antibracket $(\cdot,\cdot)$. We let $S$ denote the master action defined 
on this BV configuration space $\manM$.  This master action $S$ is required to
be of ghost number zero: $\gh{S}=0$, and will of course classically satisfy 
the Master Equation $(S,S)=0$. We also let $\Gamma^A$ denote local 
coordinates on $\manM$ (in Darboux coordinates $\Gamma^A$ are just the 
fields $\phi$ and antifields $\phi^*$).  In terms of local 
coordinates the antibracket is determined by the odd
Poisson bivector $E^{AB}=(\Gamma^A,\Gamma^B)$.

\noi
Let us now turn to the superfield formulation. In this case we simply 
consider one odd coordinate $\theta$, which we here take to be of 
opposite ghost number as compared with the previous section: 
$\gh{\theta}=-1$. We will consider $\theta$ as a Grassmann-odd version of
ordinary time in exactly the same manner as above, and
we thus allow $\Gamma^A$ to depend on $\theta$, $i.e.$, 
\begin{equation}
  \Gamma^A=\Gamma^A_0+\theta \Gamma^A_1\,,
\end{equation}
Obviously
\begin{equation}
\label{eq:BVghostno}
  \p{\Gamma^A_0}=\p{\Gamma^A_1}+1=\p{\Gamma^A}\,, \qquad
  \gh{\Gamma^A_0}=\gh{\Gamma^A_1}-1=\gh{\Gamma^A}\,,
\end{equation}
and the path space will thus have an {\em even} symplectic 
structure (see Sec.~\ref{sec:geometry}). The corresponding Poisson bracket
is given in coordinates by
\begin{equation}
\begin{array}{rcl}
\label{eq:induced-PB}
  \{\Gamma^A_0,\Gamma^B_0\} &=& 0 \cr
  \{\Gamma^A_0,\Gamma^B_1\} &=& E^{AB}(\Gamma_0) \cr
\{\Gamma^A_1,\Gamma^B_1\} &=& \Gamma^C_1
\partial_C E^{AB}(\Gamma_0) \,.
\end{array}
\end{equation}
This Poisson bracket obviously carries zero ghost number. Now we define the 
quantity
\begin{equation}
  \label{eq:BV-Omega}
\Omega(\Gamma_0,\Gamma_1) ~\equiv~ \int \! d \theta~ S(\Gamma(\theta))\,,
\end{equation}
By construction $\gh{\Omega}=1$,
and we note that $\Omega$ will be nilpotent:
\begin{equation}
  \{\Omega,\Omega\} = 0\,.
\label{Omeganil}
\end{equation}
In fact, this nilpotency condition is here to be viewed as a Poisson-bracket
Master Equation. But it immediately raises the question: Can this $\Omega$ 
also be formally considered as the BRST charge corresponding to a Hamiltonian
system with constraints? Although we do not allow the fields $\Gamma^A$ to
depend in addition on a {\em new} bosonic coordinate ``time'', 
this is in fact the case.

\noi
It is not difficult
to understand the nature of the associated Hamiltonian system of constraints.
Let $S$ be the 
extended master action of a gauge theory described by an initial action 
$S_0(q^i)$ and gauge generators $R^i_\alpha$ which we for simplicity take
to be linearly independent (the discussion can be easily generalized
to the reducible case).  They form a possibly open algebra
\begin{equation}
  [R_\alpha,R_\beta]=C^\gamma_{\alpha \beta}R_\gamma+\ldots\,.
\end{equation}
where dots means the terms vanishing on the stationary surface
of the action $S_0$.  Thus in the BV formulation we need, for the
minimal sector, the fields of the initial theory $q^i,\, \gh{q^i}=0$
(which we for simplicity take to be bosonic),
ghosts $c^\alpha,\,\gh{c^\alpha}=1$, and all their antifields.  As
usual, we combine fields into $\phi^A$, and  antifields into
$\phi^*_A$. The BV antibracket and ghost number assignments 
are
\begin{equation}
(\phi^A,\phi^*_B) = \delta^A_B\,, \qquad
\gh{\phi^*_A}=-\gh{\phi^A}-1\,.
\end{equation}
The master action constructed according to the BV prescription to satisfy the
classical Master Equation $(S,S)=0$ is then
\begin{equation}
  \label{eq:BVmaster}
S=S_0 + q^*_i R^i_\alpha c^{\alpha} - \half c^*_\gamma C^\gamma_{\alpha\beta}
c^\alpha c^\beta+\ldots\,.
\end{equation}
where the dots denotes possible terms of higher order in antifields.

\noi
We now allow $\phi^A,\phi^*_A$ to depend on $\theta$.  The
expansion of $\phi,\phi^*$ in $\theta$ thus reads 
\begin{equation}
q^i=q_0^i-\theta \gamma^i\,, \qquad
q^*_i=\pi_i+\theta p_i\,, \qquad
c^\alpha=c_0^\alpha +\theta \eta^\alpha \,, \qquad
c^*_\alpha=\rho_\alpha +\theta \cP_\alpha\,. \qquad
\end{equation}
Moreover, it follows from eq. \eqref{eq:BVghostno}
that ghost number assignments are:
\begin{equation}
\begin{array}{c}
\gh{q_0^i}=\gh{p_i}=0\,,\quad \gh{c^\alpha}=\gh{\eta^\alpha}=1\,, \quad
\gh{\cP_\alpha}=\gh{\pi_i}=-1\,, \cr
\gh{\eta^\alpha}=2\,,\qquad \gh{\rho_\alpha}=-2\,.
\end{array}
\end{equation}

\noi
The Poisson bracket (\ref{eq:induced-PB}) becomes
explicitly 
\begin{equation}
\begin{array}{c}
\{q_0^i,p_j\}=\delta^i_j\,,\qquad
\{\gamma^i,\pi_j\}=\delta^i_j\,,\cr
\{c_0^\alpha,\cP_\beta\}=\delta^\alpha_\beta\,,\qquad
\{\eta^\alpha,\rho_\beta\}=\delta^\alpha_\beta\,,
\end{array}
\end{equation}
Substituting (\ref{eq:BVmaster}) in (\ref{eq:BV-Omega})
and integrating over $\theta$ we arrive at
\begin{equation}
  \label{eq:Omega-explicit}
\Omega ~=~ -\gamma^i \partial_i S_0+
p_i R^i_\alpha c_0^\alpha-
\pi_i R^i_\alpha \eta^\alpha+
\pi_i(\gamma^j \partial_j R^i_\alpha)c_0^\alpha-
\half \cP_\gamma C^\gamma_{\alpha \beta} c_0^\alpha c_0^\beta
-\rho_\alpha C^\gamma_{\alpha\beta}\eta^\alpha c_0^\beta+\ldots\,,
\end{equation}
where dots denote higher order terms in the variables $\cP,\pi$ and $\rho$.
We will see that they are to be identified with
ghost momenta.  Eq. \eqref{eq:Omega-explicit} can formally be identified
with the BRST charge of a system with constraints. Using the ghost
number assignments it is easy to see that $\gamma^i$ and $c_0^\alpha$
are the ghosts associated with first class constraints $\partial_i S$
and $p_iR^i_{\alpha}$, and $\pi_i\,, \cP_\alpha$ are their conjugate
momenta.  The variables $\eta^\alpha$ and $\rho_\alpha$ are simply the
ghosts for ghosts and their momenta associated with the reducible
constraints $\partial_iS$. To be precise, the Lagrangian gauge
generators $R^i_\alpha$ are the reducibility functions for
the constraints $\partial_i S_0$ due to the Noether identity
$R^i_\alpha \partial_i S_0=0$. The corresponding term $\pi_i
R^i_\alpha \eta^\alpha$ indeed enters (\ref{eq:Omega-explicit}).
It should be emphasized that all these identifications are in an algebraic 
sense only: There is no analogue of the ordinary time coordinate of the 
Hamiltonian system.

\noi
An interesting open question concerns the role of quantum corrections to the
master action $S$. Suppose we expand the solution to
the full quantum Master Equation 
\begin{equation}
\frac{1}{2} (S,S) ~=~ i\hbar \Delta S
\end{equation}
in powers of $\hbar$, and insert this
full solution into the definition (\ref{eq:BV-Omega}). The nilpotency condition
(\ref{Omeganil}) will then be broken by $\hbar$-corrections on the right
hand side. What is the interpretation
of the $\hbar$-corrections to the BRST charge $\Omega$? Perhaps this is related
to canonical quantization of
the Poisson bracket and the corresponding Hamiltonian quantum Master
Equation $[\hat{\Omega}(\hbar),\hat{\Omega}(\hbar)]=0$. 
In any case, the question deserves a more detailed study.

\section{Geometry of the Super Path Space}\label{sec:geometry}

\noi
It is useful to clarify the geometrical meaning
of the structures entering the above superspace formulations, and view
them in greater generality. In particular, it is instructive to see how the
antibracket and the usual Poisson bracket enter on equal footing.
In this section, which will be a bit more abstract, we find it convenient to 
even use the same symbol for the two, namely $[\cdot,\cdot]_\manM$. 
One must of course keep in mind that the odd and even brackets have odd and 
even Grassmann parities, respectively. Now let
$\manM$ be a symplectic manifold (which can be even or odd), and let
$[\cdot,\cdot]_\manM$ be the corresponding Poisson bracket or antibracket,
depending on the Grassmann parity. We denote by $n$
the dimension of $\manM$ and $\kappa$
the Grassmann parity of the bracket $[\cdot,\cdot]_\manM$, $i.e.$,
\begin{equation}
\p{\pg{f}{g}_\manM}=\p{f}+\p{g}+\kappa\,,
\end{equation}
The exchange relation, the Leibniz rule and the Jacobi identity
are then neatly summarized, for both brackets, by
\begin{equation}
\begin{array}{rcl}
\label{eq:Jacobi}
\pg{f}{g}_\manM&=& -(-1)^{(\p{f}+\kappa)(\p{g}+\kappa)}\pg{g}{f}_\manM\, \cr
\pg{f}{gh}_\manM&=&\pg{f}{g}_\manM
h+(-1)^{(\p{f}+\kappa))\p{g}}g\pg{f}{h}_\manM\, \cr
\pg{f}{\pg{g}{h}_\manM}_\manM &=&
\pg{\pg{f}{g}_\manM}{h}_\manM+\pg{g}{\pg{f}{h}_\manM}_\manM
(-1)^{(\p{f}+\kappa)(\p{g}+\kappa)}\,.
\end{array}
\end{equation}
for any functions $f,g$ and $h$ on $\manM$.  In local coordinates
$\Gamma^A$ on $\manM$ we write generically
$E^{AB}=\pg{\Gamma^A}{\Gamma^B}_\manM$ for both kinds of brackets.

\noi
Let in addition $\Sigma$ be a supermanifold
of dimension $k$ and of coordinates $x^i$. 
We assume for simplicity that it is compact. Let there in addition be a 
volume form $d\mu(x)=\rho(x)dx=\rho(x)dx^1\ldots dx^k$ on $\Sigma$.
We denote by $\cE$ the super-path space, i.e. the space of smooth maps from
$\Sigma$ to $\manM$.  In local coordinates each map is described
by the functions $\Gamma^A(x)$.  As $\manM$ is symplectic, and
$\Sigma$ has the above volume form, then~\cite{[AKSZ]} the
super path space $\cE$ is also symplectic (see also section 4.3 of ref.
\cite{super} and ref. \cite{B}).  Indeed, for any functionals $F,G$ we define
\begin{equation}
\label{eq:path-bracket}
\pg{F}{G}_\cE=(-1)^{(\p{F}+\p{d\mu})\p{d\mu}}
\int d\mu(x)
(F\vardr{\Gamma^A(x)} E^{AB}(\Gamma(x))
\vardl{\Gamma^B(x)} G)\,.
\end{equation}
Here we have made use of the following conventions for the functional
derivatives: for infinitesimal variation
$\delta \Gamma^A(x)$ we write
\begin{equation}
\delta F[\Gamma]~=~
\int d \mu (x) \delta \Gamma^A(x) (\vardl{\Gamma^A(x)}F[\Gamma])~=~
\int (F[\Gamma] \vardr{\Gamma^A(x)}) \delta \Gamma^A(x)d \mu (x)\,.
\end{equation}
In particular, left and right derivatives are then related by
\begin{equation}
  \vardl{\Gamma^A(x)}F~=~
(-1)^{\p{d\mu}+(\p{F}+\p{d \mu}+1)(\p{\Gamma^A}+\p{d \mu})}
F \vardr{\Gamma^A(x)}\,,
\end{equation}
where $\p{d \mu }$ is the Grassmann parity of the measure.  
Note that the Grassmann parity of the functional derivative 
$\vardl{\Gamma^A(x)}$ is $\p{\Gamma^A}+\p{d \mu}$.

\noi
Let us first state some obvious properties of the bracket
structure~\eqref{eq:path-bracket}.  First,
the Grassmann parity $k^\prime$ of the bracket~\eqref{eq:path-bracket}
is related to the Grassmann parity of the bracket $\pg{\cdot}{\cdot}_\manM$
by $\kappa^\prime=\kappa+\p{d\mu}$.  The bracket
\eqref{eq:path-bracket} obviously satisfies ~\eqref{eq:Jacobi}
with $\kappa$ being the $\kappa^\prime$ and is thus a
Poisson bracket or an antibracket, depending on its Grassmann parity.  
Taking $F$ and $G$ in \eqref{eq:path-bracket} to be
\begin{equation}
  \label{eq:association}
F[\Gamma]=\int d\mu(x) f(\Gamma(x))\,, \qquad
G[\Gamma]=\int d\mu(x) g(\Gamma(x))\,,
\end{equation}
for some functions $f,g$ on $\manM$ we arrive at
\begin{equation}
\pg{F}{G}_\cE=\int d \mu \pg{f}{g}_\manM\,.
\end{equation}

\noi
Let there in addition be given a vector field
$q=q^i\dl{x^i}$ on $\Sigma$.  We assume that $div_{d \mu}(q)=0$, which implies
that $\int\! d\mu~ q f=0$ for any function $f$ on
$\Sigma$. The vector field $q$ can be lifted to a
vector field $Q$ on the super path space
$\cE$~\cite{[AKSZ]}. In coordinates we have for any functional
$F[\Gamma]$
\begin{equation}
QF[\Gamma]=\int d \mu(x)
(( q^i(x) \dl{x^i} \Gamma^A(x)) \frac{\delta}{\delta
  \Gamma^A(x)}F[\Gamma])\,.
\end{equation}
An important observation is that $Q$ is
an Hamiltonian vector field with respect to the bracket
~\eqref{eq:path-bracket}.  Indeed,
let $V_A(\Gamma)$ be the symplectic
potential on $\manM$; for the symplectic 2-form we have
\begin{equation}
E_{AB}=(\partial_A V_B -
(-1)^{\p{A}\p{B}}
\partial_B V_A )
(-1)^{\p{B}(\kappa+1)}\,,
\end{equation}
where $\kappa$ is the parity of the symplectic form.  Then for an
arbitrary  functional $F[\Gamma]$ we have
\begin{equation}
  QF=-\pg{C}{F}_\cE\,, \qquad
  C=\int\,d\mu(x)\, (q\, \Gamma^A(x))\,V_A(\Gamma(x))\,,
\end{equation}
with $\p{C} = \p{d\mu}+\p{q} + \kappa$.
Note that if $q$ is an odd nilpotent vector field on $\Sigma$,
then the corresponding Hamiltonian $C$ automatically satisfies the 
classical Master Equation $\pg{C}{C}_\cE=0$.  Another important
property of $Q$ is that for any functional $F$
of the form~\eqref{eq:association} we have $QF=-\pg{C}{F}_\cE=0$.

\noi
These properties of the super path space bracket
allows one to directly construct a BV master action
\begin{equation}
\label{eq:form}
  W = \alpha C + \beta F \,,
\end{equation}
for some functional $F$ from ~\eqref{eq:association} and any $f$ satisfying
$\pg{f}{f}_\manM=0$. This holds for arbitrary coefficients $\alpha$ and 
$\beta$.  When the Grassmann parity of this $W$ is odd, it simply has
the interpretation as a BRST-like charge $\Omega$. It was
shown in ref~\cite{[AKSZ]} that the
BV master actions corresponding to Chern-Simons theory and $2D$ topological
sigma models have precisely the same structure. In all
these cases one chooses $\Sigma$ to be an odd tangent
bundle over the even manifold $\Sigma_0$, with the odd nilpotent vector 
field $q$ being the De Rham differential
on $\Sigma_0$. Remarkably,
the BV master action of the 2D Poisson sigma model used
in~\cite{[Felder]} for the construction of the
Kontsevich star product~\cite{[Kontsevich]} also has 
just the form \eqref{eq:form}. 

\noi
Surprising relations between the Poisson brackets of 
Hamiltonian BFV quantization and antibrackets of Lagrangian quantization
have recently been discovered for topological gauge theories in a quite 
different manner \cite{G} (see also refs. \cite{ND,GST,BM}).
It would also be interesting to consider
the isomorphism between the Poisson bracket and the antibracket \cite{BH}
in the light of this superfield construction.  
 
\noi
Finally, let us explicitly make contact to the examples we 
gave in the previous sections. For the first case we choose $\Sigma$
to be a $(1|1)$ supermanifold with coordinates $t$ and $\theta$. We
also choose $\manM$ to be an even symplectic manifold, and simply take
as measure $d\mu=dt\, d\theta$, as well as an odd operation $\der=\theta
\frac{d}{dt}$. A smooth map $\Sigma \to \manM$ is given by the set of
functions $\Gamma^A(t,\theta)=\Gamma_0^A+\theta \Gamma_1^A$. Using
the representation
\begin{equation}
  \Gamma_0^A(t)=\int d \theta\, \theta \,\Gamma^A(t,\theta)\,, \qquad
  \Gamma_1^A(t)=\int d \theta \, \Gamma^A(t,\theta)\,,
\end{equation}
and explicitly integrating over $\theta$
in the definition~\eqref{eq:path-bracket} we
arrive at
\begin{equation}
\begin{array}{rcl}
\label{eq:AB-explicit}
\ab{\Gamma^A_0(t)}{\Gamma^B_0(t^\prime)} &=& 0\, \cr
\ab{\Gamma^A_0(t)}{\Gamma^B_1(t^\prime)} &=& \omega^{AB}(\Gamma_0)
\delta(t-t^\prime)\, \cr
\ab{\Gamma^A_1(t)}{\Gamma^B_1(t^\prime)} &=&
\Gamma^C_1\partial_C
 \omega^{AB}(\Gamma_0) \delta(t-t^\prime)\,.
\end{array}
\end{equation}
In its turn the odd nilpotent vector field $D$, considered as acting
on functionals, is a Hamiltonian vector field with Hamiltonian 
\begin{equation}
  C=\int\, dt\, d\theta\, V_A \der \Gamma^A\,.
\end{equation}
Choosing Darboux coordinates $p,q$ on $\manM$ one arrives at
the standard form $\int dt p^0_{i}{\dot{q}}_{0}^i$. Thus we
see that the Hamiltonian action~\eqref{S} has precisely
the ``geometrical'' form~\eqref{eq:form} with $f$
being the super BRST charge $\Omega+\theta H$. The only
difference is that $f$ in this case explicitly depends
on $\theta$.  

\noi
In the case of the inverse construction of Sec.~\ref{sec:invers} one chooses
$\Sigma$ to be a one dimensional space with Grassmann-odd coordinate $\theta$
and $\manM$ as an antisymplectic manifold. Using the general
formula~\eqref{eq:path-bracket}
one arrives directly at the explicit form of the odd path space
Poisson bracket~\eqref{eq:induced-PB}. In this case we simply take $\alpha=0$
and $F$ to be the master action $S$ in eq. \eqref{eq:form}.
We have thus shown how both of these
cases follow directly from the above general framework.

\vspace{1cm}
\noindent
{\sc Acknowledgement:}~ The work of M.A.G. is partially supported by
INTAS grant YSF-98-156, RFBR grant 98-01-01155, Russian Federation
President Grant 99-15-96037, and the Landau Scholarship Foundation,
Forschungszentrum J\"ulich. We thank I.A.~Batalin, K.~Bering,
A.M.~Semikhatov, I.Yu.~Tipunin and especially I.V.~Tyutin for fruitful
discussions.

 \end{document}